\documentclass[twocolumn]{aastex63}
\usepackage{amsmath}
\usepackage{here}
\usepackage{graphics}
\usepackage[T1]{fontenc}

\newcommand{\vecb}[1]{\mbox{\boldmath $#1$}}

\newcommand{\rhog}{\rho_{\mathrm{g}}}
\newcommand{\rhod}{\rho_{\mathrm{d}}}
\newcommand{\rhogup}{\rho_{\mathrm{g},0}}
\newcommand{\rhodup}{\rho_{\mathrm{d},0}}
\newcommand{\gas}{\mathrm{g}}
\newcommand{\dst}{\mathrm{d}}
\newcommand{\drift}{\mathrm{drift}}

\newcommand{\sigmad}{\Sigma_{\dst}}

\newcommand{\tstop}{t_{\mathrm{stop}}}

\newcommand{\taus}{\mathrm{St}}

\newcommand{\cs}{c_{\mathrm{s}}}

\newcommand{\hd}{H_{\mathrm{d}}}
\newcommand{\vk}{v_{\mathrm{K}}}

\newcommand{\dri}{\mathrm{drift}}

\newcommand{\rhoint}{\rho_{\mathrm{int}}}

\newcommand{\epmid}{\epsilon_{\mathrm{mid}}}
\newcommand{\epmidini}{\epsilon_{\mathrm{mid},0}}

\newcommand{\coag}{\mathrm{coag}}

\newcommand{\epcgro}{\epsilon_{\mathrm{c,gro}}}
\newcommand{\epcpla}{\epsilon_{\mathrm{c,plts}}}
\newcommand{\tauscpla}{\mathrm{St}_{\mathrm{c,plts}}}
\newcommand{\tausini}{\mathrm{St}_{\mathrm{0}}}
\newcommand{\zc}{z_{\mathrm{c}}}

\received{--}
\accepted{--}

\shorttitle{Dust coagulation assisted by streaming instability}
\shortauthors{Tominaga and Tanaka}

\begin{document}

\title{Dust coagulation assisted by streaming instability in protoplanetary disks}

\correspondingauthor{Ryosuke T. Tominaga}
\email{tominaga.r.aa@m.titech.ac.jp}
\author[0000-0002-8596-3505]{Ryosuke T. Tominaga}
\affiliation{Department of Earth and Planetary Sciences, Institute of Science Tokyo, 2-12-1, Ookayama, Meguro, Tokyo 152-8551, Japan}

\author[0000-0001-9659-658X]{Hidekazu Tanaka}
\affiliation{Astronomical Institute, Graduate School of Science, Tohoku University, 6-3, Aramaki, Aoba-ku, Sendai 980-8578, Japan}



\begin{abstract}
The streaming instability is a promising mechanism for planetesimal formation. The instability can rapidly form dense clumps that collapse self-gravitationally, which is efficient for large dust grains with the Stokes number on the order of 0.1. However, dust growth models predict that collisional fragmentation prevents dust grains from growing to such sizes. We perform local simulations of the streaming instability and measure characteristic collision velocities and collision rates of dust grains based on their trajectories in moderate clumping. The collision velocities are on the order of 0.1 percent of the sound speed or lower, implying that dust grains can overcome the fragmentation barrier via the clumping. We also find that the collision rates are appreciably high regardless of the low collision velocities. Corresponding timescales are on the order of ten Keplerian periods or shorter, suggesting that dust grains can overcome the drift barrier as well. This streaming-instability-assisted (SI-assisted) coagulation greatly relaxes the conditions for planetesimal formation as recently implied.
\end{abstract}



\section{Introduction}\label{sec:intro}

Formation of planetesimals from dust grains is one of the key steps in plant formation. There are some obstacles known to prevent dust coagulation in protoplanetary disks. One of the major obstacles is the radial drift of dust toward the central star \citep[e.g.,][]{Whipple1972,Weidenschilling1977}, which is caused by different orbital velocities of dust and gas. While dust grains tend to move at Keplerian velocities, gas moves at sub-Keplerian velocities because of a pressure gradient force in the radial direction. Because of the headwind of gas, the dust grains lose angular momentum and move inward. The maximum drift speed is on the order of 0.01 au/yr in protoplanetary disks. Such fast drifting motion prevents dust grains from growing larger before they reach the central star \citep[e.g.,][]{Brauer2008}.

Another obstacle is collisional fragmentation \citep[e.g.,][]{Blum1993}. Experiments and numerical simulations suggested critical velocities beyond which dust grains are disrupted \citep[e.g.,][]{Blum2000,Blum2008,Wada2009}. Critical velocities depend on material properties, including chemical composition and monomer size of dust aggregates \citep[e.g.,][]{Chokshi1993,Dominik1997}. According to \citet{Wada2009}, the critical velocity for water ice aggregates with 0.1 $\mu\mathrm{m}$-sized monomers is $\sim50\;\mathrm{m/s}$, which is comparable to the maximum drift speed. Recent studies have shown that $\mathrm{CO}_2$ ice is much less sticky than water ice \citep[][]{Musiolik2016a,Musiolik2016b}. They suggested that the critical velocity for $\mathrm{CO}_2$ ice is about 10 times lower than that for water ice. Thus, fragmentation may prevent dust coagulation even in a cold outer region when dust is covered by $\mathrm{CO}_2$ ice mantles \citep[see also][]{Pinilla2017,Okuzumi2019}.

One of the promising mechanisms to circumvent the obstacles is hydrodynamic instabilities \citep[e.g.,][and references therein]{Simon2022,Lesur2023}. Among them, the streaming instability (SI) is the leading and extensively studied mechanism \citep[e.g.,][]{YoudinGoodman2005}. The SI is driven by aerodynamic interaction between drifting dust and gas \citep[e.g.,][]{Squire2018b,Squire2020,Magnan2024} and can increase local dust densities by orders of magnitude \citep[e.g.,][]{Johansen2007,Bai2010b}. Numerical simulations of the SI have shown that the local dust density can exceed the Roche density, meaning that resulting dust clumps gravitationally collapse into planetesimals \citep[e.g.,][]{Johansen2007nature,Johansen2009b,Simon2016,Schafer2017,Abod2019}. In contrast to the gravitational instability \citep[][]{Safronov1972,Sekiya1983}, the SI does not require self-gravity and is thus expected to operate at various stages of the disk evolution in terms of dust mass.

Conditions for planetesimal formation via the SI have been investigated in previous studies \citep[][]{Carrera2015,Yang2017,Li2021,Lim2024}. They performed vertically stratified simulations for multiple parameter sets and derived critical dust-to-gas ratios as a function of the Stokes number, $\taus$, which is the dimensionless stopping time of dust. \citet{Carrera2015} found the critical dust-to-gas surface density ratio to be $\simeq0.02$ for $\taus\simeq0.1$ and larger for smaller $\taus$. Although the critical values have been updated in the subsequent studies, they also show that the critical dust-to-gas ratio decreases as $\taus$ increases to $\taus\sim0.1$ \citep[][]{Yang2017,Li2021,Lim2025}. \citet{Lim2024} simulated the SI with forced gas turbulence in stratified simulations. They found that the condition is well described in terms of midplane dust-to-gas ratios, and the critical value is $\simeq1$ for $\taus=0.1$ and a few times higher for $\taus=0.01$ (see Figure 4 therein). The critical values are similar among the runs with different strength of the turbulence. These studies indicate that the SI can form planetesimals if dust grains grow to the size of $\taus\sim0.01-0.1$. The required $\taus$ may be even smaller since \citet{Yang2017} and \citet{Li2021} observed strong clumping in runs with $\taus=10^{-3}$.

The amount of concentrated dust grains via the SI depends on $\taus$. \citet[][]{Johansen2007} showed cumulative dust density distributions obtained from their unstratified simulations (see their Figure 11). They showed that more dust grains reside in overdense regions for $\taus=1$ than for $\taus=0.1$. According to the stratified simulations with multiple dust species \citep[][]{Bai2010b}, the mass fraction of dust grains in overdense regions seems largest for $\taus\sim10^{-0.5}$. The concentration of smaller dust grains ($\taus<0.1$) is limited, and the mass fraction of dust grains in dense regions can be an order of magnitude smaller than for $\taus\sim0.1-1$. Recent high-resolution simulations with $\taus=0.01$ also showed that the dust densities exceed the Roche density only at limited grid cells \citep[][]{Lim2025}.

Besides, the previous numerical simulations have shown that it takes long time for the dust density to reach the Roche density when $\taus$ is small. According to \citet{Li2021}, duration time of this pre-clumping phase can be a few hundreds of the Keplerian periods for $\taus\sim10^{-2}$ unless dust-to-gas surface density ratios are initially high enough. This would be problematic for planetesimal formation in local regions such as a gas bump. \citet{Carrera2022b} simulated the SI in a gas bump and showed that the clumping with $\taus\simeq0.01$ is so slow that the dust densities do not exceed the Roche density before the dust crosses the bump \citep[see also][]{Carrera2022}.

These studies suggest that planetesimal formation via the SI requires sufficiently large $\taus$ (e.g., 0.1-0.3) from a viewpoint of planetesimal formation efficiency. The required dust size seems comparable to the drift-limited size \citep[e.g., see][]{Brauer2008}. The dust growth model predicts much smaller $\taus$ when the critical velocity is lower than $\sim10\;\mathrm{m/s}$ and fragmentation hinders dust growth \citep[e.g., see][]{Drazkowska2023}. As mentioned above, dust grains will experience such disruptive collisions even in a cold region because of the low critical velocity for $\mathrm{CO}_2$ ice. Therefore, an additional mechanism to circumvent the fragmentation barrier is necessary to trigger the strong clumping and planetesimal formation via the SI.

The dust growth beyond the fragmentation barrier potentially occurs during moderate clumping due to the SI. \citet[][]{Klahr2020} suggested that large-scale gas turbulence is ineffective in stirring dust grains once they are locally concentrated. In such a case, the SI-driven turbulent motion will be the dominant source of collision velocities. Previous studies of the SI showed that the collision velocities can be so low that dust grains avoid disruptive collisions in overdense regions \citep[][]{Johansen2007supp,Balsara2009,Bai2010b}. According to \citet{Schreiber2018}, velocity dispersion of dust grains in the SI turbulence decreases with increasing dust density \citep[see also][]{Bai2010b}. Based on these studies, we estimated coagulation timescales in our previous study and suggested that dust grains have enough time to grow before the dust densities exceed the Roche density via the SI \citep[][]{TominagaTanaka2023}. The above process is thus a promising way to bridge the gap: collision velocities start decreasing once the SI develops, which allows dust grains to grow beyond the fragmentation barrier. 

In this work, we perform numerical simulations of the SI to measure coagulation rates more precisely. Our measurement takes into account time variations in dust densities and velocities along their trajectories in the SI turbulence. We demonstrate that the SI-assisted coagulation greatly relaxes the conditions for planetesimal formation.

We note that \citet{Ho2024} performed numerical simulations incorporating both the SI and dust growth. They also imply that the dust growth during the SI can relax the conditions for planetesimal formation since the Stokes number is enhanced. In their study, they simply adopt a constant strength of turbulence to mimic the SI-induced collisions. In contrast, we focus on revealing collision properties of the SI-assisted coagulation, based on which we develop a simple model to show relaxed conditions. Therefore, this study and \citet{Ho2024} are complementary to each other.

This paper is organized as follows. We describe basic equations and numerical methods in Section \ref{sec:method} and show the results of the measurement of coagulation rates as well as characteristic dust densities and collision velocities. We discuss implications to planetesimal formation and validities of our simulations in Section \ref{sec:discussion}. We conclude in Section \ref{sec:conclusion}.

\section{Method}\label{sec:method}
We use the Athena code \citep{Stone2008,Stone2010,Bai2010a} to simulate the SI in an unmagnetized isothermal gas. Athena is a grid-based code and includes a super-particle module developed by \citet{Bai2010a} for modeling a dusty gas system. The super-particle module allows us to measure physical properties of (sub-)grid-scale dust collisions (collision velocities and rates), which is in contrast to fully grid-based codes based on two-fluid approximation of dusty gas disks.

We adopt the local shearing coordinates $(x,\;y,\;z)=(r-R,\;R(\phi-\Omega t), z)$, where $(r,\;\phi,\; z)$ denotes the cylindrical coordinates, $R$ denotes radial distance from the central star of mass $M_{\ast}$, and $\Omega$ is the Keplerian angular velocity \citep[][]{Goldreich1965b}. We use the cornered transport upwind (CTU) integrator \citep[][]{Colella1990} and the HLLC Rieman solver with the piecewise parabolic reconstruction to calculate numerical fluxes \citep[see][]{Colella1984,Stone2008}. We assume an axisymmetric disk and perform two-dimensional simulations. In this work, we omit the vertical stellar gravity. The following subsections describe basic equations, numerical settings and diagnostics to estimate coagulation rates in detail.

\subsection{Basic equations}\label{subsec:basiceqs}

The equations for the gas in the local shearing coordinates are
\begin{equation}
\frac{\partial\rhog}{\partial t}+\nabla\cdot\left(\rhog \vecb{u}\right)=0,\label{eq:gas_eoc}
\end{equation}
\begin{align}
\frac{\partial\rhog\vecb{u}}{\partial t} &+\nabla\cdot\left(\rhog\vecb{u}\vecb{u}+P\vecb{I}\right)\notag\\
&=\rhog\left[2\vecb{u}\times\Omega\vecb{e}_z+3\Omega^2x\vecb{e}_x+\frac{\rhod}{\rhog}\frac{\vecb{v}-\vecb{u}}{\tstop}\right].\label{eq:gas_mom}
\end{align}
In the above equations, $\rhog$ and $\vecb{u}$ denote the gas density and velocities in the local frame, and $P$ denotes the gas pressure. We use the isothermal equation of state, $P=\rhog\cs^2$, where $\cs$ is the isothermal sound speed. Unit vectors in each direction are denoted by $\vecb{e}_x,\;\vecb{e}_y$ and $\vecb{e}_z$. The final term on the right hand side of Equation (\ref{eq:gas_mom}) represents the backreaction from dust to gas. The backreaction depends on the dust density $\rhod$ and the stopping time $\tstop$. 

The super-particle module of Athena utilizes the following equation of motion \citep[][]{Bai2010a}:
\begin{equation}
\frac{d\vecb{v}_i}{dt} = -2\eta \vk\Omega\vecb{e}_x+2\vecb{v}_i\times\Omega\vecb{e}_z+3\Omega^2x_i\vecb{e}_x-\frac{\vecb{v}_i-\vecb{u}}{\tstop},\label{eq:sp_eom}
\end{equation}
where the subscript $i$ represents a label of a super-particle. As described in \citet[][]{Bai2010a}, the orbital motion of super-particles is forced to be super-Keplerian by a constant force term, $-2\eta\vk\Omega\vecb{e}_x$, which mimics the effect of the radially outward pressure gradient in the gas disk. The coefficient $\eta\vk$ represents the velocity difference from the Keplerian velocity $\vk$ that the gas should have in a dust-free case, where
\begin{equation}
\eta\equiv-\frac{1}{2}\left(\frac{\cs}{\vk}\right)^2\frac{d\ln\rhog}{d\ln r}.
\end{equation}
In this work, we assume $\eta\vk=0.05\cs$. We refer readers to \citet[][]{Bai2010a} for more details of this approach. 

We employ the triangular-shaped cloud (TSC) interpolation scheme to calculate the drag force on super-particles. This scheme is also used to calculate the backreaction from super-particles to gas at grid points. 

\subsection{Numerical settings}\label{subsec:num_set}

Our study is concerned with efficiency of dust coagulation especially for small dust grains whose Stokes number is much less than unity. Since the spatial scale of the SI is shorter for smaller $\taus$, we set the radial and vertical extents ($L_x,\;L_z$) of the domain to be $0.01H$ in this work. We fix the domain size in a parameter study where $\taus$ ranges from 0.01 to 1. We consider this setting reasonable although the most unstable wavelength is much larger for $\taus=1$. This is because such large dust grains should sediment in a disk, and the dust scale height $\hd$ can be on the order of $10^{-2}H$ \citep[][]{Dubrulle1995,YL2007}: 
\begin{equation}
\hd=\sqrt{\frac{\alpha}{\alpha+\taus}}H\sim10^{-2}H\left(\frac{\alpha}{10^{-4}}\right)^{1/2}\left(\frac{\taus}{1}\right)^{-1/2},\label{eq:hd}
\end{equation}
where $\alpha$ is the dimensionless measure of turbulence strength \citep[][]{Shakura1973}. Recent ALMA observations suggest weak turbulence of $\alpha\sim10^{-5}-10^{-3}$ \citep[][]{Flaherty2015,Flaherty2017,Pinte2016,Villenave2022}, which motivates us to use $10^{-4}$ as the reference value in Equation (\ref{eq:hd}). Modes whose length scale is longer than $\hd$ would not develop, and thus we expect that adopting the small domain size for runs with $\taus=1$ would be valid. We adopt the periodic boundary condition in all directions \citep[e.g.,][]{Balbus1991}, which seems valid if the domain size is smaller than the dust scale height.

The number of cells is $N_x\times N_z=256\times256$, and the total number of super-particles is $9 N_x N_z$. \citet[][]{Bai2010a} suggested that one particle per cell on average is sufficient for convergence. Our simulations use the larger number of particles to estimate their velocity dispersion as in \citet{Schreiber2018}. The Courant-Friedrichs-Lewy number was fixed to 0.5.

Initial conditions are the same as those in previous studies \citep[e.g.,][]{Bai2010a,Schreiber2018,Baronett2024}. The initial gas density $\rhogup$ is uniform. The super-particles are randomly distributed, and the resulting density fluctuations are seed perturbations. We initialize the radial and azimuthal velocities using the equilibrium drift velocities with the uniform dust-to-gas ratio $\epsilon\equiv\rhodup/\rhogup$ \citep[][]{Nakagawa1986}. Initial vertical velocities of dust and gas are set to be zero.

We set $\rhogup=\cs=\Omega=1$ in code units. Parameters in this study are the Stokes number $\taus\equiv\tstop\Omega$ and the initial dust-to-gas ratio $\epsilon$. Table \ref{tab:params} summarizes parameter sets explored in this study and times ($t_{\mathrm{s}}, \;t_{\mathrm{e}}$) between which we measure coagulation rates.

\begin{deluxetable}{c|cccc|c}[tp]
\tablecaption{Sets of parameters and measured coagulation rates} \label{tab:params}
\tablehead{
\colhead{Label} &
\colhead{$\epsilon$\tablenotemark{a}}&
\colhead{$\taus$} & 
\colhead{$t_{\mathrm{s}}\Omega$\tablenotemark{b}} &
\colhead{$t_{\mathrm{e}}\Omega$\tablenotemark{b}} &
\colhead{$(t_{\coag}\Omega)^{-1}$} 
}
\startdata
mr0.3st1 &
0.3 & 0.01 & 60 & 120 & $9.2\times 10^{-3}$ \\ 
mr0.3st3 &
0.3 & 0.03 & 60 & 120 & $2.9\times 10^{-2}$ \\ 
mr0.3st10 &
0.3 & 0.1 & 60 & 120 & $2.8\times 10^{-2}$ \\ 
mr0.5st1 &
0.5 & 0.01 & 60 & 120 & $1.7\times 10^{-2}$ \\ 
mr0.5st3 &
0.5 & 0.03 & 60 & 120 & $4.7\times 10^{-2}$ \\ 
mr0.5st10 &
0.5 & 0.1 & 60 & 120 & $3.3\times 10^{-2}$ \\ 
mr0.5st30 &
0.5 & 0.3 & 100 & 200 & $7.8\times 10^{-3}$ \\ 
mr1st1 &
1.0 & 0.01 & 60 & 120 & $3.0\times 10^{-2}$ \\ 
mr1st3 &
1.0 & 0.03 & 60 & 120 & $6.8\times 10^{-2}$ \\ 
mr1st10 &
1.0 & 0.1 & 60 & 120 & $4.0\times 10^{-2}$ \\ 
mr1st30 &
1.0 & 0.3 & 60 & 120 & $2.1\times 10^{-2}$ \\ 
mr1st100 &
1.0 & 1.0 & 100 & 200 & $1.1\times 10^{-4}$ \\ 
mr3st1 &
3.0 & 0.01 & 60 & 120 & $5.1\times 10^{-2}$ \\ 
mr3st3 &
3.0 & 0.03 & 60 & 120 & $1.1\times 10^{-1}$ \\ 
mr3st10 &
3.0 & 0.1 & 60 & 120 & $8.1\times 10^{-2}$ \\ 
mr3st30 &
3.0 & 0.3 & 60 & 120 & $3.7\times 10^{-2}$ \\ 
mr3st100 &
3.0 & 1.0 & 60 & 120 & $3.4\times 10^{-3}$ \\ 
mr10st1 &
10.0 & 0.01 & 60 & 120 & $5.4\times 10^{-2}$ \\ 
mr10st3 &
10.0 & 0.03 & 60 & 120 & $1.3\times 10^{-1}$ \\ 
mr10st10 &
10.0 & 0.1 & 60 & 120 & $1.3\times 10^{-1}$ \\ 
mr10st30 &
10.0 & 0.3 & 60 & 120 & $6.3\times 10^{-2}$ \\ 
mr10st100 &
10.0 & 1.0 & 60 & 120 & $1.7\times 10^{-2}$ \\ 
\enddata
\tablenotetext{a}{Initial dust-to-gas ratio}
\tablenotetext{b}{Times between which we analyze the data to measure coagulation rates and velocity dispersions (see Section \ref{subsec:diagnostics}).}
\end{deluxetable}

\subsection{Diagnostics}\label{subsec:diagnostics}

\begin{figure*}
\begin{center}
	\hspace{100pt}\raisebox{20pt}{
	\includegraphics[width=2\columnwidth]{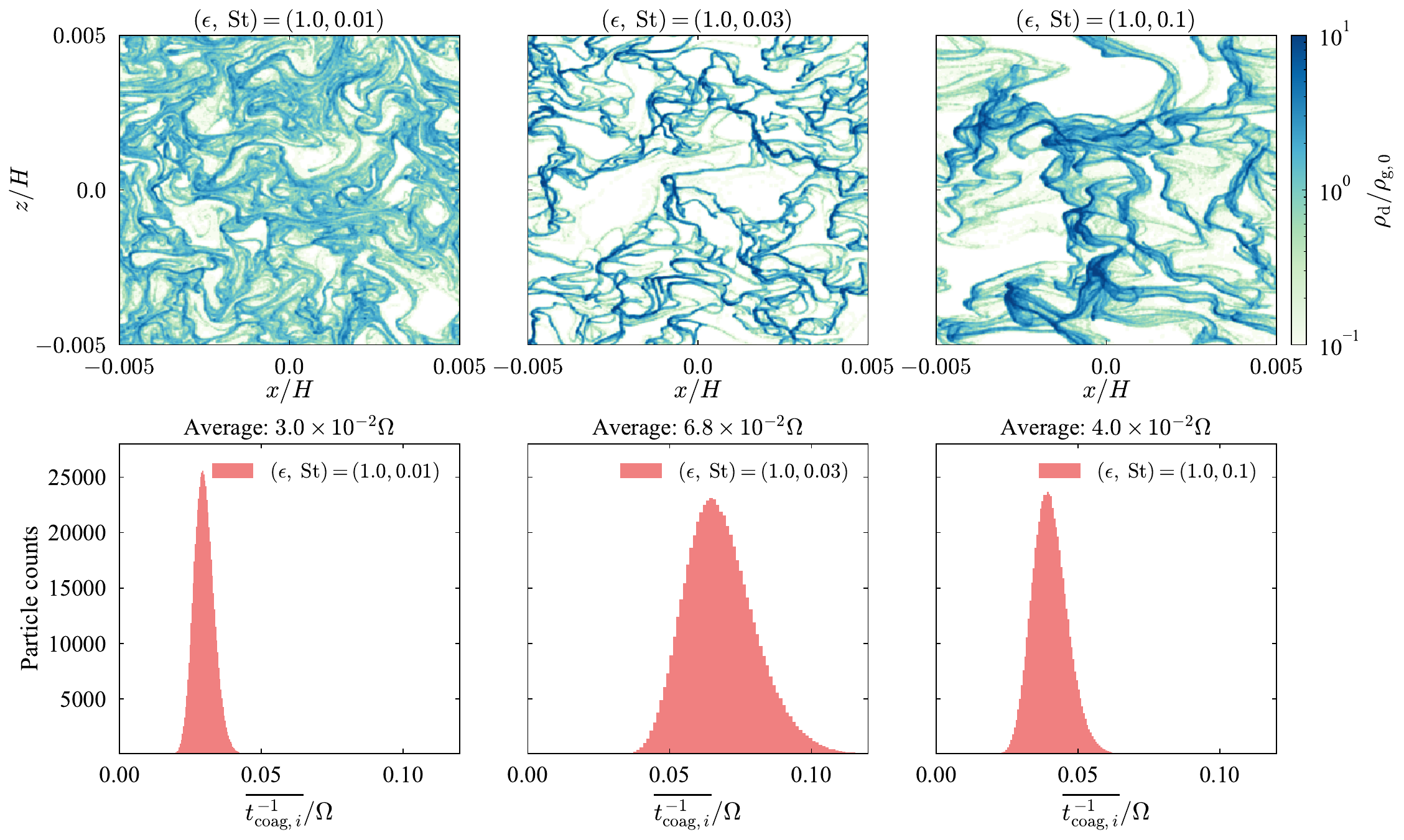} 
	}
	\end{center}
	\vspace{-30pt}
\caption{Top three panels show dust density distributions at $t=t_{\mathrm{s}}$ for the mr1St1, mr1st3, and mr1st10 runs. Bottom panels show histograms of the time-averaged coagulation rates $\overline{t_{\coag,i}^{-1}}$ (see Equation (\ref{eq:tcoag_ave})). The histograms show single-peak profiles, with which we can estimate the typical evolutionary timescales of the systems.}\label{fig:rhod_col_num}
\end{figure*}

Collision probability depends on dust density and collision velocity, which vary spatially and temporally in the SI-driven turbulence. In this study, we take into account their variations along particle trajectories and derive statistically-averaged collision rates. The collision rates are estimated for equal-mass collisions since we assume single-species dust in each run.

A collision rate at the $i$th particle position is given by
\begin{align}
t_{\mathrm{coll},i}^{-1} &= \frac{\rho_{\dst,i}4\pi a^2\Delta v_{i}}{m},\label{eq:coll}
\end{align}
where $a$ and $m$ denote the size and the mass of a dust grain, and $\Delta v_i$ denotes the collision velocity. We assume dust grains in the Epstein regime to relate $\taus$ and the dust size:
\begin{equation}
\taus =\sqrt{ \frac{\pi}{8}}\frac{\rhoint a}{\rhog\cs}\Omega,\label{eq:epstein}
\end{equation}
where $\rhoint \equiv 3m/4\pi a^3$ is the internal density. We define the coagulation timescale $t_{\coag,i}$ as the size-doubling timescale, 
\begin{equation}
t_{\coag,i}^{-1}=\frac{1}{a}\frac{da}{dt}=\frac{t_{\mathrm{coll},i}^{-1}}{3},
\end{equation}
where we assume perfect sticking \citep[cf.][]{TominagaTanaka2023}. Using the above equations, we obtain
\begin{equation}
t_{\coag,i}^{-1}=\sqrt{\frac{\pi}{8}}\frac{\rho_{\dst,i}\Delta v_{i}}{\rho_{\gas}\cs\taus}\Omega.\label{eq:tcoag_i_def}
\end{equation}
 Since the gas motion induced by the SI is almost incompressible \citep[e.g., see][]{Youdin2007,Johansen2007}, $\rho_{\gas}$ in the above equations is assumed to be the unperturbed gas density $\rhogup$. 

Collision velocities are different between particle pairs. In this study, we assume that the collision velocity for the $i$th particle is represented by velocity dispersion of super-particles in a cell where the $i$th particle is located. The velocity dispersion $\sigma(t, x_i, z_i)$ is given by
\begin{equation}
\sigma(t, x_i,z_i)^2=\frac{1}{n_{\mathrm{p},i}}\sum^{n_{\mathrm{p},i}}\left(\vecb{v}'_i-\left<\vecb{v}_i'\right>_{\mathrm{cell}}\right)^2,
\end{equation}
where $n_{\mathrm{p},i}$ represents the number of particles that are located in the same cell as the $i$th particle, $\vecb{v}'_i$ is the velocity relative to the Keplerian velocity, and $\left<\vecb{v}'_i\right>_{\mathrm{cell}}$ is the average velocity of these particles. 

To obtain average coagulation rates, we first integrate $t_{\mathrm{coag},i}^{-1}$ and estimate the number of collisions that the $i$th particle experiences for $t_{\mathrm{s}}\leq t\leq t_{\mathrm{e}}$. The time-averaged coagulation rate for the $i$th particle is given by
\begin{equation}
\overline{t_{\coag,i}^{-1}}\equiv\frac{1}{\Delta T_{\mathrm{s,e}}}\int^{t_{\mathrm{e}}}_{t_{\mathrm{s}}}t_{\coag,i}^{-1}dt',\label{eq:tcoag_ave}
\end{equation}
where $\Delta T_{\mathrm{s,e}}\equiv t_{\mathrm{e}}-t_{\mathrm{s}}$. We output particle data every $0.5\Omega^{-1}$ and use them for the time integration. We take the average of the coagulation rates for all super-particles:
\begin{equation}
t_{\coag}^{-1} \equiv \frac{1}{N_{\mathrm{p}}}\sum_i\overline{t_{\coag,i}^{-1}}.\label{eq:tcoag_ave_all}
\end{equation}
This gives the evolutionary timescale of the system due to coagulation.

We also calculate $t_{\coag}^{-1}$-weighted (coagulation-rate-weighted) average values of the dust density and the velocity dispersion:
\begin{equation}
\overline{\rho_{\dst,i}}\equiv\frac{\int^{t_{\mathrm{e}}}_{t_{\mathrm{s}}}t_{\coag,i}^{-1} \rhod dt'}{\Delta T_{\mathrm{s,e}}\overline{t_{\coag,i}^{-1}}},
\end{equation}
\begin{equation}
\overline{\sigma_i}\equiv\frac{\int^{t_{\mathrm{e}}}_{t_{\mathrm{s}}}t_{\coag,i}^{-1} \sigma dt'}{\Delta T_{\mathrm{s,e}}\overline{t_{\coag,i}^{-1}}}.
\end{equation}
Their ensemble average is given as follows:
\begin{equation}
\left<\rhod\right>\equiv\frac{1}{N_{\mathrm{p}}}\sum_i\overline{\rho_{\dst,i}}, \label{eq:w_ave_dens}
\end{equation}
\begin{equation}
\left<\sigma\right>\equiv\frac{1}{N_{\mathrm{p}}}\sum_i\overline{\sigma_i}.\label{eq:w_ave_vdisp}
\end{equation}
These represent characteristic dust density and velocity dispersion for the SI-assisted coagulation.

\section{Results}\label{sec:results}

\begin{figure}[tp]
	\begin{center}
	\hspace{100pt}\raisebox{20pt}{
	\includegraphics[width=\columnwidth]{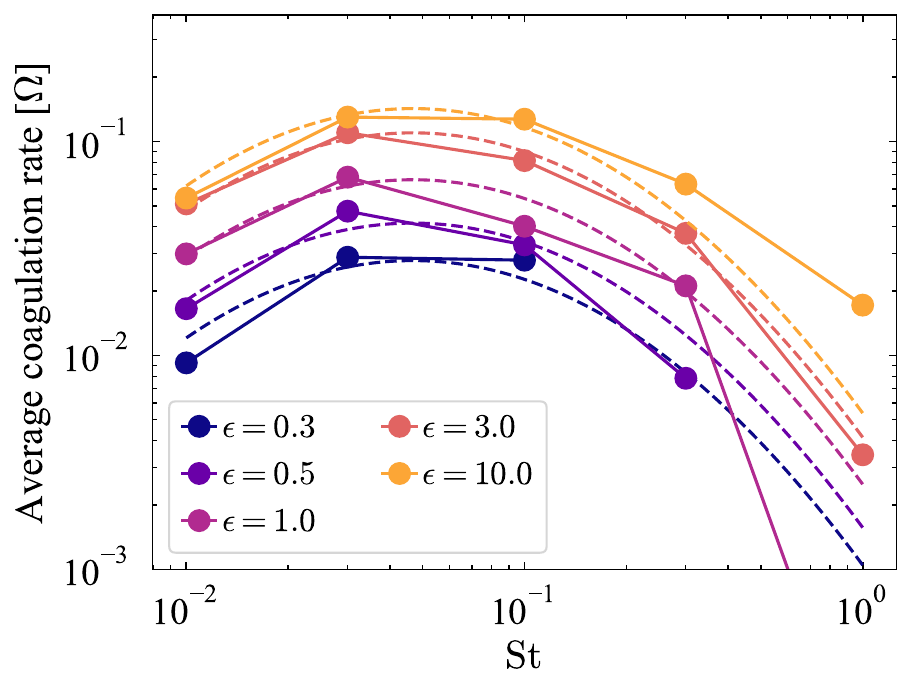} 
	}
	\end{center}
	\vspace{-30pt}
\caption{Coagulation rates averaged for all super-particles (see Equation (\ref{eq:tcoag_ave_all})) as a function of $\taus$. Color shows initial dust-to-gas ratios. Dashed lines represent the approximated formula (Equation (\ref{eq:tcoag_fit_func})), with which we aim to approximate the coagulation rates for $\taus\leq0.1-0.3$. We note that the coagulation rate of the mr1st100 run is $\sim10^{-4}\Omega$ (see Table \ref{tab:params}).} 
\label{fig:tcoag_ave_all_depend}
\end{figure}

The top panels of Figure \ref{fig:rhod_col_num} show dust density distributions at $t=t_{\mathrm{s}}$ from three runs (mr1st1, mr1st3, mr1st10), where we find that the maximum density is about $10\rhogup$ or larger. Spatial scales of the clumping vary with $\taus$, which is the typical trend of the SI. We refer readers to previous studies for basic features of the instability \citep[e.g.,][]{Johansen2007}.

In the bottom panels of Figure \ref{fig:rhod_col_num}, we show histograms of the time-averaged coagulation rates $\overline{t_{\coag,i}^{-1}}$ from each run (see Equation (\ref{eq:tcoag_ave})). The histograms show single peak profiles, which is also the case for the other runs. In this case, the average coagulation rate is expected to well represent the typical evolutionary timescale of the system. In Figure \ref{fig:tcoag_ave_all_depend}, we show the average coagulation rates (Equation (\ref{eq:tcoag_ave_all})) as a function of $\taus$ for different initial dust-to-gas ratios $\epsilon$. The measured coagulation rates are also listed in Table \ref{tab:params}. The average coagulation rate of the mr1st100 run is on the order of $10^{-4}\Omega$. Most of the other runs show coagulation rates of $\sim10^{-2}\Omega - 10^{-1}\Omega$.

The dashed lines in Figure \ref{fig:tcoag_ave_all_depend} represent an approximated function of $(t_{\coag}\Omega)^{-1}$:
\begin{equation}
(t_{\coag}\Omega)^{-1}\simeq \frac{\epsilon}{\taus} \times 10^{f(\taus)} g(\epsilon),\label{eq:tcoag_fit_func}
\end{equation}
\begin{equation}
f(\taus) = A_2(\log\taus)^2 + A_1\log\taus + A_0,
\end{equation}
\begin{equation}
g(\epsilon)=\frac{B_0}{B_1 + \epsilon},
\end{equation}
where $A_2=-0.80,\;A_1= -1.14,\;A_0= -2.46,\; B_1=1.47$, and $B_0= 1.78$. We derive these coefficients by fitting a part of the data points in Figure \ref{fig:tcoag_ave_all_depend}: the points of $\taus\leq0.1$ for $f(\taus)$, and those of $\epsilon\geq0.5$ and $\taus\leq0.3$ for $g(\epsilon)$. The data of $\taus=1$ is not used for fitting since our focus is the potential growth of small dust grains.


\begin{figure*}
\plottwo{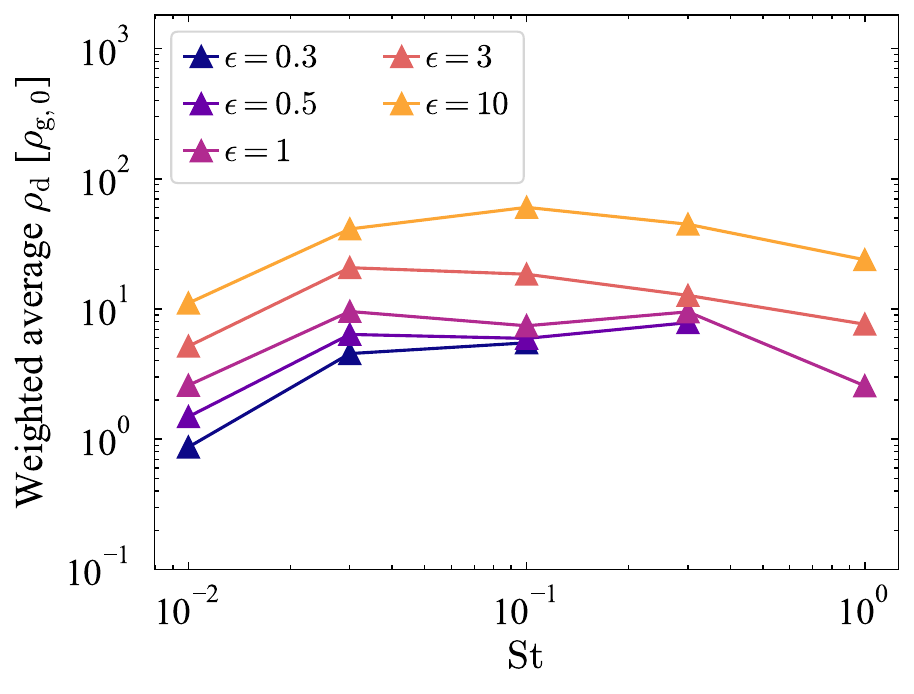}{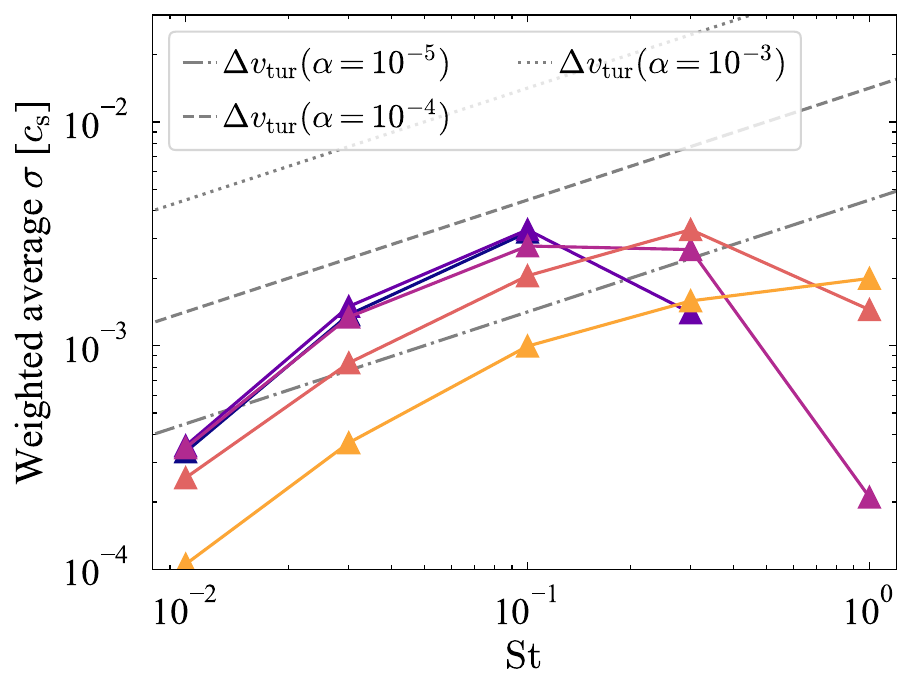}
\caption{Average dust density (left) and velocity dispersion (right) as a function of $\taus$ (see Equations (\ref{eq:w_ave_dens}) and (\ref{eq:w_ave_vdisp})). Those are weighted by the coagulation rates and thus represent the characteristic densities and velocities relevant to the coagulation. We note that the dust density on the left panel is normalized by the initial gas density. The density increase is larger in smaller $\epsilon$. On the right panel, we also show the collision velocities driven by gas turbulence $\Delta v_{\mathrm{tur}}=\sqrt{2\alpha\taus}\cs$. The SI-induced collision velocities are comparable to $\Delta v_{\mathrm{tur}}(\alpha=10^{-5})$. As found in the previous studies \citep[e.g.,][]{Bai2010b,Schreiber2018}, we found that the weighted average velocity dispersion decreases as $\epsilon$ increases beyond unity. \label{fig:weighted_average}}
\end{figure*}

The coagulation rates are maximized at $\taus\simeq0.03-0.1$. To provide an interpretation of this result, we plot the coagulation-rate-weighted average $\rhod$ and $\sigma$ in Figure \ref{fig:weighted_average}. For a given $\epsilon$, the weighted average density is largest at $0.03\leq\taus\leq0.3$\footnote{The unstratified simulations of the previous studies showed that the SI with $\taus=1$ enhances the dust density more than with $\taus=0.1$ \citep[e.g.,][]{Johansen2007,Johansen2014}. This is not seen in our simulations, which might be due to the fixed domain sizes (see Section \ref{subsec:num_set} for its validation).}. Among the runs, $\left<\rhod\right>/\rho_{\dst,0}$ is largest in the mr0.3st10 run. The velocity dispersion increases by a factor of $\simeq10$ as $\taus$ increases from 0.01 to 0.1. The increase becomes modest for $\taus>0.1$, and $\left<\sigma\right>$ eventually decreases except for $\epsilon=10$. Therefore, the increase in the coagulation rates seen for $\taus\leq0.1$ is due to the increase in both $\left<\rhod\right>$ and $\left<\sigma\right>$. The decrease in the coagulation rate for $\taus\geq0.1$ is mainly due to the decrease in the cross-section of dust ($t_{\mathrm{coag}}^{-1}\propto\taus^{-1}\Omega$, see Equation (\ref{eq:tcoag_i_def})). 

The grey dashed lines on the left panel of Figure \ref{fig:tcoag_ave_all_depend} show turbulence-induced velocities for equal-mass collisions, $\Delta v_{\mathrm{tur}}=\sqrt{2\alpha\taus}\cs$ \citep[][]{Ormel2007}. We find that $\left<\sigma\right>$ is comparable to the turbulence-induced velocity for $\alpha=10^{-5}$. It is worth noting that $\left<\sigma\right>$ decreases as $\epsilon$ when $\epsilon$ exceeds unity, which is especially seen for $\taus\leq0.1$. The similar trend was found in \citet{Schreiber2018} although the dust-to-gas ratio at which $\left<\sigma\right>$ starts decreasing is smaller than they showed for $\sigma$ (e.g., see blue contours of Figures 11 and 21 therein).

\begin{figure}[tp]
	\begin{center}
	\hspace{100pt}\raisebox{20pt}{
	\includegraphics[width=\columnwidth]{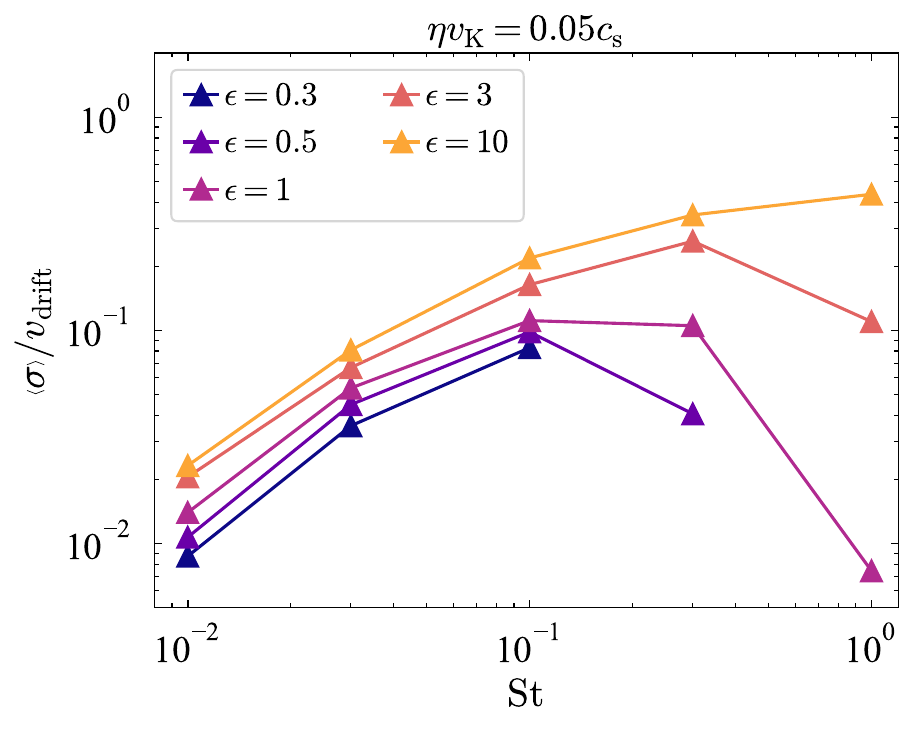} 
	}
	\end{center}
	\vspace{-30pt}
\caption{Ratios of the measured velocity dispersion and the drift speed, which include both radial and azimuthal components (Equation (\ref{eq:drift_all})). We adopt the initial $\epsilon$ to estimate the drift speeds although dust grains in overdense regions move more slowly.} 
\label{fig:sigma_vs_vdrift}
\end{figure}

We also compare the measured velocity dispersion with the drift speed $v_{\dri}$ \citep[][]{Nakagawa1986}:
\begin{equation}
v_{\dri}\equiv\sqrt{v_{\dri,r}^2+v_{\dri,\phi}^2}, \label{eq:drift_all}
\end{equation}
\begin{equation}
v_{\dri,r}=-\frac{2\taus}{(1+\epsilon)^2+\taus^2}\eta\vk,
\end{equation}
\begin{equation}
v_{\dri,\phi}=-\frac{1+\epsilon}{(1+\epsilon)^2+\taus^2}\eta\vk.
\end{equation}
Figure \ref{fig:sigma_vs_vdrift} shows $\left<\sigma\right>/v_{\dri}$ as a function of $\taus$. We find that the velocity dispersion is lower than the drift speed by a factor of $10^{-2}-10^{-1}$ for $\epsilon\leq 1$. We note that the azimuthal drift speed $|v_{\dri,\phi}|$ is $\simeq\eta\vk/(1+\epsilon)$ for $\taus\ll1$ and greater than the radial drift speed. Thus, the denominator (i.e., $v_{\dri}$) is almost constant for small $\taus$, because of which the increasing trend of $\left<\sigma\right>/v_{\dri}$ is similar to one on the right panel of Figure \ref{fig:weighted_average}. The velocity dispersion for $\taus<0.1$ is comparable to $v_{\dri,r}$ for $\epsilon\lesssim1$ and 10 times larger for $\epsilon=10$.

\begin{figure}[tp]
	\begin{center}
	\hspace{100pt}\raisebox{20pt}{
	\includegraphics[width=\columnwidth]{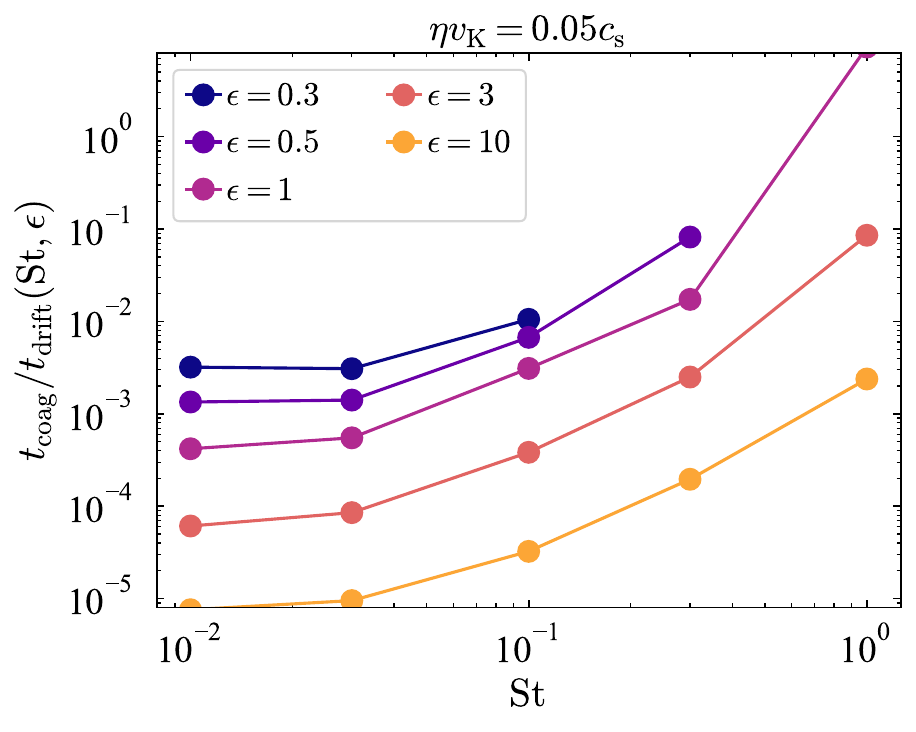} 
	}
	\end{center}
	\vspace{-30pt}
\caption{Ratios of the coagulation timescale and the radial drift timescale given by Equation (\ref{eq:t_drift}). We use the initial dust-to-gas ratios to evaluate the drift timescale. Except for the mr1st100 run, the coagulation timescale is much shorter than the drift timescale. Thus, the radial drift will not limit the dust growth during the clumping.} 
\label{fig:tdrift_vs_tcoag}
\end{figure}

Figure \ref{fig:tdrift_vs_tcoag} compares the average coagulation timescale $t_{\coag}$ with the radial drift timescale $t_{\dri}$:
\begin{equation}
t_{\dri}(\taus,\epsilon)=\frac{r}{|v_{\dri,r}|}=\frac{(1+\epsilon)^2+\taus^2}{2\taus\eta}\Omega^{-1},\label{eq:t_drift}
\end{equation}
We find that the coagulation for $\taus\leq0.3$ is faster than the radial drift. The drift timescale will be much longer since the local dust density is increased by the clumping, and the backreaction reduces the drift speed \citep[e.g.,][]{Bai2010b}. Even for $\taus=1$, the coagulation timescale is shorter than the drift timescale when the initial dust-to-gas ratio is larger than unity. Thus, dust grains grow during the clumping before they reach the central star.

For $\taus\leq0.1$, the coagulation timescale is shorter than the duration time of pre-clumping phase reported in \citet[][]{Li2021}. The midplane dust-to-gas ratio in their simulations is $\simeq0.3-3.5$ for $0.01\leq\taus\leq1$ (see Figure 4 therein), and the duration time is longer than $10^2\Omega^{-1}$ for $\taus\leq0.1$. Our simulations with $\epsilon=0.3-3$ show that the coagulation timescale ranges from $9\Omega^{-1}$(mr3st3) to $109\Omega^{-1}$ (mr0.3st1). The coagulation timescales from the mr0.5st30 and mr1st30 runs are $\simeq128\Omega^{-1}$ and $48\Omega^{-1}$, which are longer than the pre-clumping periods for $\taus=0.3$ and $\epsilon>0.5$ \citep[see Table 2 in][]{Li2021}. Therefore, we expect that the SI-assisted coagulation will increase $\taus$ up to $\simeq0.3$ in the pre-clumping phase.

\section{Discussion}\label{sec:discussion}

\subsection{SI-assisted coagulation in a gas bump}

Previous studies investigated the SI in a gas bump where dust grains are enriched \citep[e.g.,][]{Taki2016,Auffinger2018,Carrera2021}. \citet{Carrera2022} showed that the SI with $\taus\simeq0.01$ cannot form planetesimals because dust grains cross the bump before the dust density exceeds the Roche density. In this case, we should compare the coagulation timescale with the crossing timescale. If we assume the radial scale of the bump to be $\sim H$, the crossing timescale is on the order of $t_{\drift}H/r\sim 5\times10^{-2}t_{\drift}$ for $\eta\vk=0.05\cs$. According to Figure \ref{fig:tdrift_vs_tcoag}, the coagulation timescale is shorter than $\sim 4\times10^{-3}t_{\drift}$ for $\taus=0.01$. This indicates that the coagulation timescale is also shorter than the crossing time. Although a spatial variation in $\eta\vk/\cs$ is not taken into account here, we expect that such small dust grains will grow during the moderate clumping, which would enable planetesimal formation in the bump.

\begin{figure}[tp]
	\begin{center}
	\hspace{100pt}\raisebox{20pt}{
	\includegraphics[width=\columnwidth]{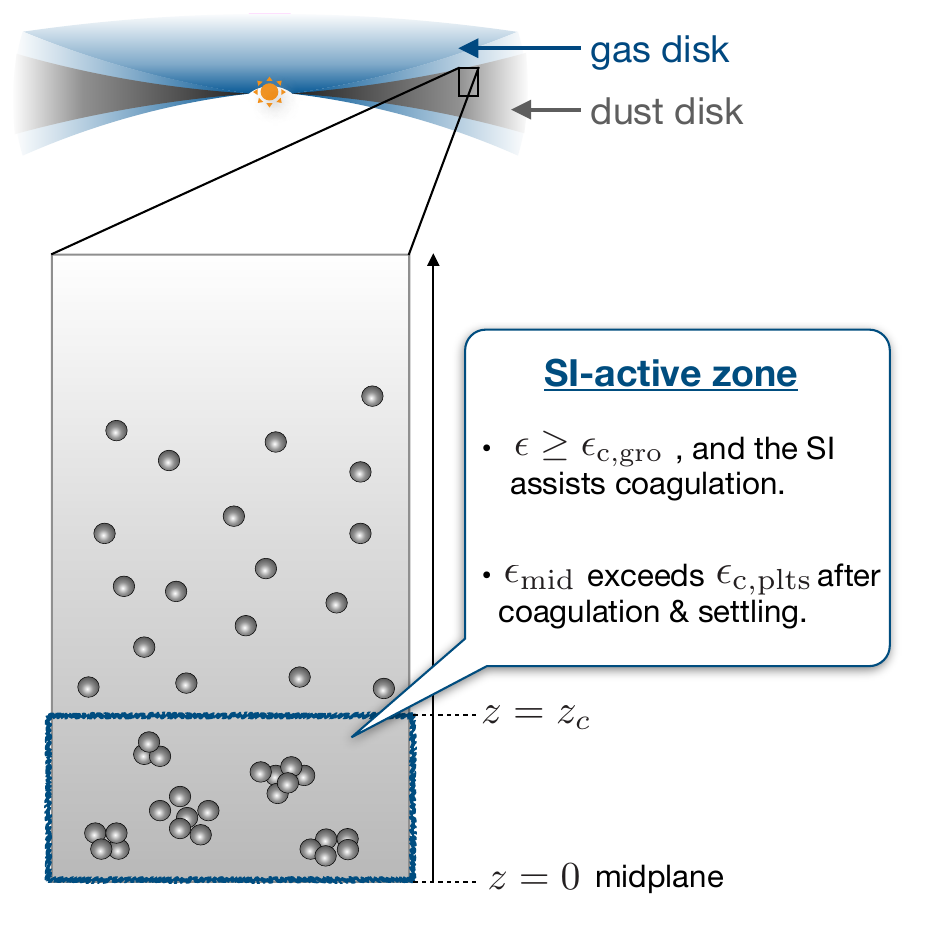} 
	}
	\end{center}
	\vspace{-30pt}
\caption{Schematic picture to show the setup with which we develop the simple model of dust growth. The SI-active zone ($|z|\leq z_{\mathrm{c}}$) is a region where the clumping due to the SI aids the dust growth. As the SI-assisted coagulation proceeds, resulting large dust grains settle to the midplane, and the midplane dust-to-gas ratio increases. The increase in $\taus$ and $\epmid$ triggers planetesimal formation.} 
\label{fig:si_active_zone}
\end{figure}

\subsection{Relaxing the conditions for planetesimal formation}

We have shown that dust grains can grow efficiently during the clumping. As suggested by \citet{Ho2024}, such dust growth relaxes the conditions for planetesimal formation. Here, we develop a simple model of dust evolution and quantitatively show how much the SI-assisted coagulation relaxes the conditions.  

Our primary focus is on dust growth beyond the barriers enabled by the SI. We thus first suppose that the dust growth is limited initially (i.e., before the onset of the SI), and dust grains of $\taus=\tausini$ are vertically distributed in a disk. We consider a range of $\tausini\geq10^{-2}$ in this work. For $\taus<\tausini$, coagulation may proceed faster than the SI because linear growth timescales of the SI are long for small dust while timescales of turbulence-induced coagulation without the clumping are independent from $\taus$ \citep[e.g.,][]{Brauer2008}. In this case, the dust growth to the size of $\tausini$ promotes the SI, which one may call ``coagulation-assisted SI". The SI-assisted coagulation concerned in this work occurs after this phase, which we discuss below.

We also suppose that, once the SI operates, the concentrated dust grains are shielded from external gas turbulence \citep[][]{Klahr2020}, and their collision velocities are determined by $\left<\sigma\right>$. The SI-assisted coagulation permits further dust growth since the collisions velocities are sufficiently low ($\lesssim10^{-3}\cs$, see the right panel of Figure \ref{fig:weighted_average}). The assumption of sticking collisions will be valid at least in an outer region beyond the water snow line, where critical velocities seem higher than a few cm/s as mentioned in Section \ref{sec:intro}. 

It is expected that the SI develops in a relatively denser region around the midplane \citep[e.g., see][and references therein]{Lesur2023}. We thus introduce a critical value of $\epsilon$ for the SI-assisted coagulation ($\epcgro$). We call a region of $\epsilon\geq\epcgro$ the SI-active zone (Figure \ref{fig:si_active_zone}). To evaluate an amount of dust grains that can grow, we assume the Gaussian distribution of the dust density. The half width of the SI-active zone $\zc$ is then given as follows: 
\begin{equation}
z_c = \hd\sqrt{2\ln\left(\frac{\epmidini}{\epcgro}\right)},\label{eq:zcrit}
\end{equation}
where the gas density is assumed to be uniform for $|z|\leq z_c$. The dust surface density of the SI-active zone $\Delta\Sigma_{\dst}$ is 
\begin{equation}
\Delta\Sigma_{\dst}=\sigmad \mathrm{erf}\left(\frac{z_c}{\sqrt{2}\hd}\right),\label{eq:dsigd}
\end{equation}
where $\sigmad$ is the total surface density, and $\mathrm{erf}(z)$ is the error function.

The dust grains in the SI-active zone settle toward the midplane as they grow. This increases the midplane dust-to-gas ratio and trigger planetesimal formation once the dust-to-gas ratio becomes sufficiently high ($\epmid\geq\epcpla$). An amount of massive clumps and thus planetesimals also depends on $\taus$. There would be a critical Stokes number $\tauscpla$, beyond which the formation rate of planetesimals dominates the coagulation rate. In the present model, we simply adopt $\tauscpla=0.3$ and assume that dust grains grow up to this size. This is motivated by the previous numerical simulations that showed the highest clumping efficiency at $\taus\simeq0.1-1$ \citep[e.g., see][]{Bai2010b}.

If all dust grains in a disk grow and settle (i.e., $\epcgro\to0$), the midplane dust-to-gas ratio increases by a factor of $\sqrt{\tauscpla/\tausini}$ (see Equation (\ref{eq:hd})):
\begin{equation}
\epmid=\epmidini\sqrt{\frac{\tauscpla}{\tausini}},\label{eq:epmid_active}
\end{equation}
where $\tausini$ and $\epmidini$ denote the Stokes number and the midplane dust-to-gas ratio before the onset of the SI, respectively. In this case, the critical value of $\epmidini$ for planetesimal formation is obtained from $\epmid=\epcpla$, or
\begin{equation}
\epmidini\sqrt{\frac{\tauscpla}{\tausini}}=\epcpla. \label{eq:simplest_epmidc}
\end{equation}
Since dust grains can grow only in the SI-active zone, we should replace $\epmidini$ with $\epmidini\Delta\sigmad/\sigmad$ in Equation (\ref{eq:simplest_epmidc}). The midplane dust-to-gas ratio after the SI-assisted coagulation is given by
\begin{equation}
\epmid=\epmidini\frac{\Delta\sigmad}{\sigmad}\sqrt{\frac{\tauscpla}{\tausini}}.
\end{equation}
We again note that we assume the Gaussian distribution for the grown dust with the scale height of $\hd\sqrt{\tausini/\tauscpla}$.\footnote{We admit that this assumption is not necessarily valid. The settling timescale $\sim(\taus\Omega)^{-1}$ can be comparable to the coagulation rate shown in Figure \ref{fig:tcoag_ave_all_depend}. We will address this issue in future work.} Using Equations (\ref{eq:zcrit}) and (\ref{eq:dsigd}), we obtain the equation for the critical value of $\epmidini$ as follows:
\begin{equation}
\epmidini\sqrt{\frac{\tauscpla}{\tausini}}\mathrm{erf}\left(\sqrt{\ln\left(\frac{\epmidini}{\epcgro}\right)}\right)=\epcpla.\label{eq:epmidc_eq}
\end{equation}
There is a real root for $\epmidini>\epcgro$.\footnote{When deriving Equation (\ref{eq:epmidc_eq}), we neglect redistribution of dust grains that are in the ``SI-dead" zone, which is the layer above the active zone. During the dust growth and settling in the SI-active zone, those dust grains of $\taus=\tausini$ should be redistributed by settling and turbulent diffusion, and some of them will reach the midplane. This means that the total dust-to-gas ratio at the midplane should be larger than one given by Equation (\ref{eq:epmid_active}). We expect that this makes little difference in our conclusion since the midplane dust-to-gas ratio is dominated by the grown dust.}

\begin{figure}[tp]
	\begin{center}
	\hspace{100pt}\raisebox{20pt}{
	\includegraphics[width=\columnwidth]{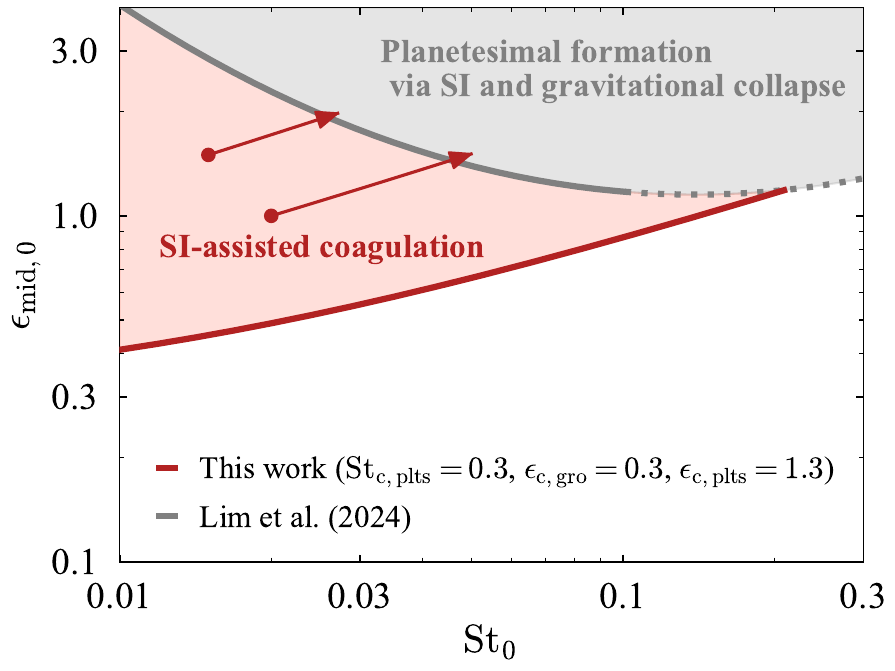} 
	}
	\end{center}
	\vspace{-30pt}
\caption{Critical values of the midplane dust-to-gas ratio as a function of $\tausini$. The red line represents the critical value derived from Equation (\ref{eq:epmidc_eq}). The assumed value of $\epcpla$ is based on the grey line that represents the critical value suggested by \citet{Lim2024}. In the grey region, the SI can form massive clumps that gravitationally collapse into planetesimals. In the red region, the SI-assisted coagulation and subsequent settling increase $\taus$ and $\epmid$ (red arrows) and enables planetesimal formation. We note that $\tausini$ in our model represents the dust size before the onset of the SI-assisted coagulation. The dust growth to this size (e.g., through collisions due to gas turbulence) is not illustrated in this figure.}
\label{fig:epmid_crit}
\end{figure}

\begin{figure*}
\gridline{\fig{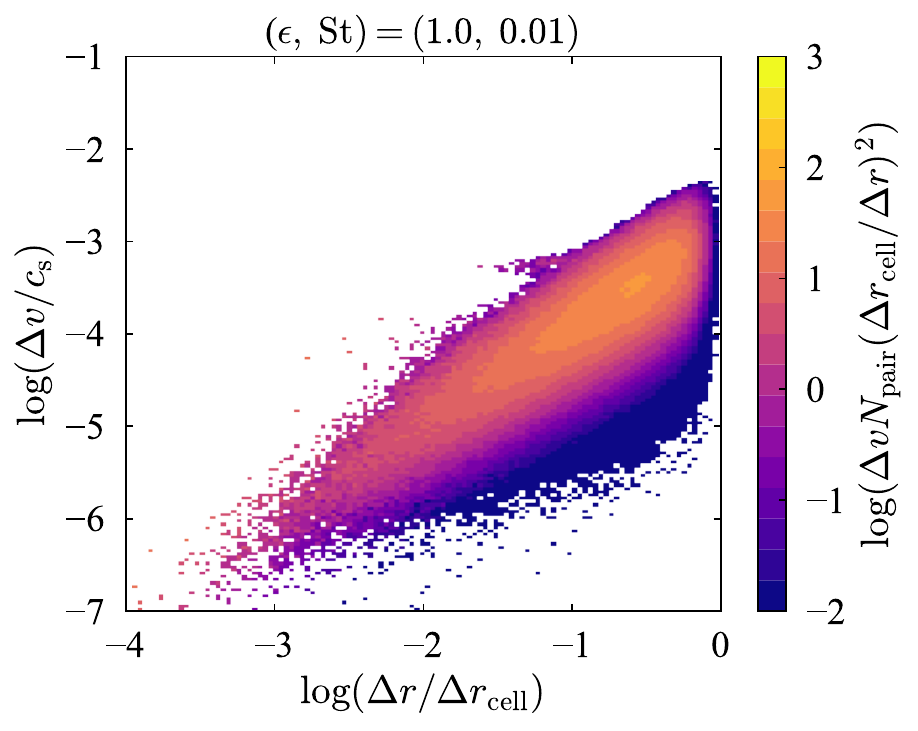}{0.3\textwidth}{}
          \fig{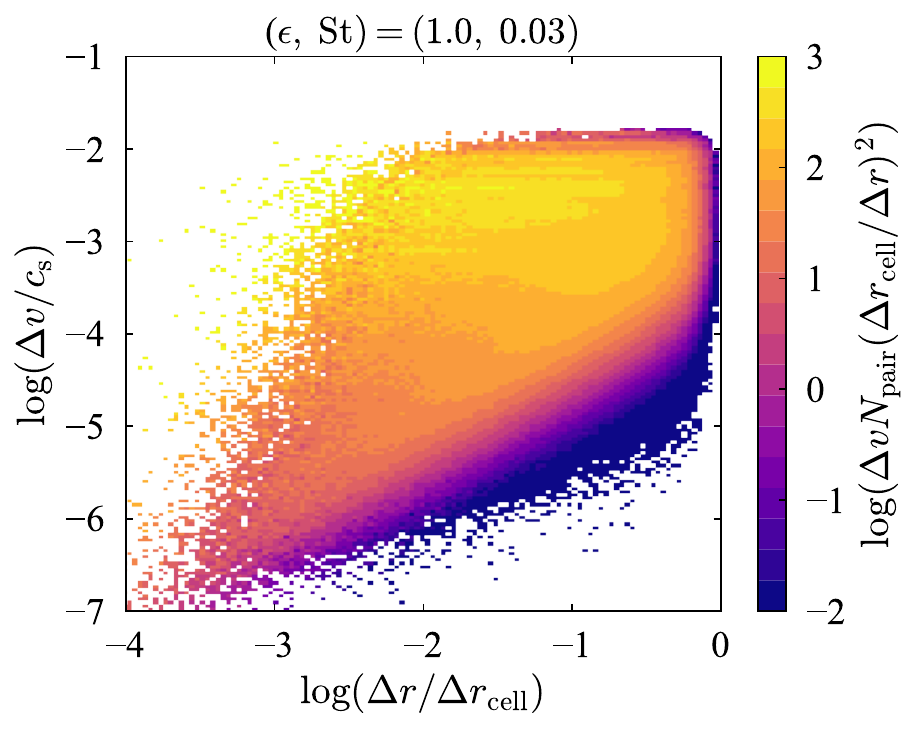}{0.3\textwidth}{}
          \fig{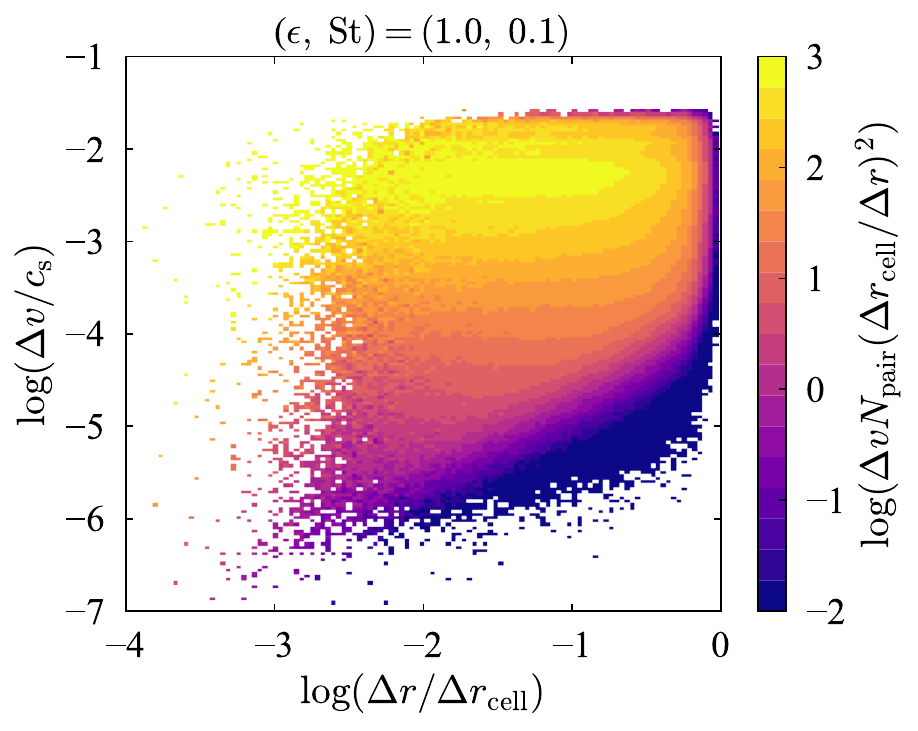}{0.3\textwidth}{}
          }
\caption{Correlation between particle distance and relative velocity in the mr1st1, mr1st3, and mr1st10 runs ($t=t_{\mathrm{s}}$). Color represents the number of super-particle pairs weighted by $\Delta v(\Delta r_{\mathrm{cell}}/\Delta r)^2$, where $\Delta r_{\mathrm{cell}}$ is the diagonal distance of a single cell. We plot such weighted values to highlight relevant regions where collision rates are high. We find relatively tight correlation in the mr1st1 run. The similar trend is found in other runs with $\taus=0.01$. The correlation becomes less tight in the runs with larger $\taus$. }\label{fig:dv_dr_cor}
\end{figure*}

Figure \ref{fig:epmid_crit} shows the critical value of $\epmidini$ for $\epcgro=0.3$ and $\epcpla\simeq1.3$. We evaluate the value of $\epcpla$ for $\taus=\tauscpla=0.3$ using the fitting formula of the critical dust-to-gas ratio given by \citet[][]{Lim2024} (Equation (16) therein), where the coagulation is not taken into account and they adopt the Toomre $Q$ value of $\simeq32$ \citep[][]{Toomre1964}. We note that \citet{Lim2024} derived the fitting formula for $\epcpla$ for $0.01\leq\taus\leq0.1$ (grey solid line). We extend the line to $\taus=0.3$ using the same formula (the grey dotted line), which should be verified in future work. If the dust size is fixed, planetesimal formation via the SI and gravitational collapse occurs only in the grey region. The SI-assisted coagulation extends the parameter space (the red region) since $\taus$ and $\epmid$ are enhanced, which are represented by red arrows. The critical value of $\epmidini$ for $\tausini=0.01$ is about ten times smaller than that for the planetesimal formation only via the SI and gravitational collapse. In this way, the SI-assisted coagulation greatly relaxes the conditions for planetesimal formation. 

We note that the degree of increase in $\rhod$ differs between the grey and red regions. In the grey region, local dust densities are enhanced by a factor of hundreds and exceed the Roche density, which is about $180\rhogup$ for $Q=32$. The enhancement factor is much lower in the red region. For example, the maximum densities are only from several to ten times higher than $\rhog$ in the run with $\taus=0.01$ and $\epsilon\leq1$. The weighted average $\rhod$ is even lower (see Figure \ref{fig:weighted_average}). Therefore, only several times increase in $\rhod$ by such moderate clumping can trigger planetesimal formation through the SI-assisted coagulation.

Previous dust evolution models adopt the critical midplane dust-to-gas ratio of unity for planetesimal formation for $\taus>10^{-2}$ \citep[e.g.,][]{Drazkowska2016,Stammler2019}. Although $\epcpla$ is higher than unity (the grey line), Figure \ref{fig:epmid_crit} validates the use of such a criteria because dust can move into the grey region via the SI-assisted coagulation. \citet{Drazkowska2016} adopted lower efficiency of planetesimal formation than that found in the stratified simulations of the SI with $\taus=0.3$ \citep[][]{Simon2016}. This is motivated by the fact that they consider planetesimal formation directly from smaller dust ($\taus<0.3$). Our model suggests that the efficiency should be as high as in the stratified simulations because the SI-assisted coagulation takes place before the planetesimal formation and enhances $\taus$.

The extent of the red region depends on $\epcpla$ as well as $\tauscpla$. It would be important to determine $\tauscpla$ by comparing the coagulation rates and the planetesimal formation rates for relatively large $\taus$ (e.g., $0.1-0.3$) in future work.

\subsection{Collision velocities on sub-grid scales}

\begin{figure}[tp]
	\begin{center}
	\hspace{100pt}\raisebox{20pt}{
	\includegraphics[width=\columnwidth]{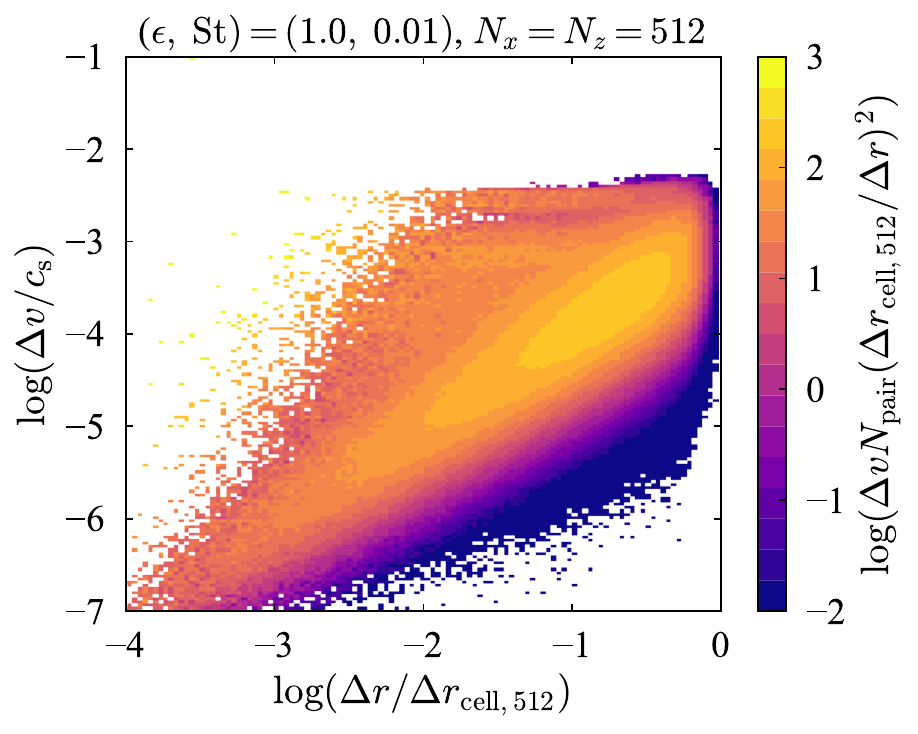} 
	}
	\end{center}
	\vspace{-30pt}
\caption{Correlation between particle distance and relative velocity in the saturated state from the mr1st1 run with $N_x=N_z=512$ ($t\Omega=60$). We note that the particle distance on the horizontal axis is normalized by $\Delta r_{\mathrm{cell},512}$, which is smaller than the normalization factor in Figure \ref{fig:dv_dr_cor}. We find that the relatively tight correlation becomes weaker in this high-resolution simulation.} 
\label{fig:dv_dr_cor_N512}
\end{figure}


We utilized the grid-scale velocity dispersion to estimate the coagulation rates. However, actual collisions occur on much smaller scales, for which the collision velocity is not necessarily represented by the grid-scale velocity dispersion \citep[see also the discussion in][]{Carrera2015}. To examine this, we measure the distances and relative velocities of super-particle pairs and investigate how the collision velocities change with the particle distance on the sub-grid scale.

Figure \ref{fig:dv_dr_cor} shows the correlation of the distance $\Delta r$ and the relative velocity $\Delta v$ of a pair of super-particles that are located in the same cell. The color scale denotes the number of super-particle pairs $N_{\mathrm{pair}}$ multiplied by $\Delta v$ and $(\Delta r/\Delta r_{\mathrm{cell}})^{-2}$, where $\Delta r_{\mathrm{cell}}$ is a diagonal distance of the cell. We multiply those factors so that the color scale highlights regions where collision rates are high.\footnote{ Because our analyses based on 2D simulations, $N_{\mathrm{pair}}\left(2\pi\Delta r\times d(\Delta r)\right)^{-1}=N_{\mathrm{pair}}\left(2\pi(\Delta r)^2d\log (\Delta r)\right)^{-1}$ represents the number density of particles in a ring whose radius and width are $\Delta r$ and $d(\Delta r)$, respectively.} For example, the m1r1 run (the left panel) shows that the collision rate is dominated by populations of $10^{-4}\lesssim\Delta v/\cs\lesssim10^{-3}$ and $0.1\lesssim\Delta r/\Delta r_{\mathrm{cell}}\lesssim 1$, where $\Delta r_{\mathrm{cell}}$ is the diagonal distance of a cell.

We find a positive correlation for $\taus=0.01$. As also shown in direct numerical simulations by \citet{Ishihara2018}, closer super-particles of $\taus\sim0.01$ have smaller relative velocities. The present correlation is due to the limited resolution. We run a simulation with $N_x=N_z=512$ and measure the correlation (Figure \ref{fig:dv_dr_cor_N512}). We find larger relative velocities especially for $10^{-3}\lesssim\Delta r/\Delta r_{\mathrm{cell},512}\lesssim10^{-1}$, where $\Delta r_{\mathrm{cell},512}$ is the diagonal length of the cell in the high-resolution run. The increase in the relative velocities might be due to turbulent motion that is not well resolved in the fiducial run \citep[e.g., see also][]{Johansen2007supp}. Nevertheless, the relative velocity is mostly below $\sim10^{-3}\cs$, and thus collisional fragmentation would be insignificant. Simulations with higher resolution are necessary for more detailed analyses. 

For larger $\taus$, the difference between relative velocities of close and distant super-particle pairs is insignificant in the runs with 256 cells (the middle and right panels). Especially on the right panel, the color map shows a horizontally extended peak around $\Delta v\sim\left<\sigma\right>$ (see also Figure \ref{fig:weighted_average}). This means that the relative velocities of close particles can be as high as those of distant particles ($\Delta r\sim\Delta r_{\mathrm{cell}}$), and those particle pairs have comparable contributions to collision rates. In these cases, the velocity dispersion well represents the actual collision velocity. We thus expect that the present results for $\taus\gtrsim0.03$ would insignificantly change in higher-resolution simulations.

\subsection{ Bouncing barrier}
 In addition to the radial drift and fragmentation barriers, the bouncing barrier is another obstacle to prevent dust growth \citep[e.g.,][]{Zsom2010,Dominik2024}. The process has been investigated from laboratory experiments \citep[e.g.,][]{Langkowski2008,Weidling2012,Kothe2013,Schrapler2022} and numerical simulations \citep[e.g.,][]{Wada2011,Schrapler2012,Seizinger2013,Arakawa2023,Oshiro2025}. The previous studies showed that relatively compact aggregates can rebound for the collision velocity lower than $1\;\mathrm{m/s}$. This indicates that the bouncing can limit dust growth before the fragmentation takes place.

 When the bouncing barrier keeps $\taus$ small, the SI becomes less efficient than often expected, while it can hold dust grains in disks for $\sim1\mathrm{Myr}$ \citep[][]{Dominik2024}; otherwise they will be depleted quickly \citep[e.g.,][]{Brauer2008,Birnstiel2009}. If the bouncing barrier limits $\taus$ to be less than $10^{-2}$, dust grains would be too tightly coupled to form a large amount of dense clumps via the SI. Overcoming the bouncing barrier is challenging, but necessary to induce such strong clumping with the help of dust coagulation for planetesimal formation. In future work, we should investigate the efficiency of the SI-assisted coagulation in the presence of the bouncing barrier.

\subsection{Caveats and Outlooks}
In this section, we describe several caveats and limitations regarding the present simulations.

\subsubsection{Unequal-mass collisions and size distributions}
Throughout this work, we have explored the coagulation rates due to equal-mass collisions. As shown in Figure \ref{fig:rhod_col_num}, the coagulation rates have a certain degree of dispersion in a single run. This means that the SI-assisted coagulation would broaden a size distribution even when dust grains are initially monodisperse. As the size distribution becomes wider, collisions due to differential drift motion would contribute to the dust growth, which is not taken into account in the present work.

We roughly estimate relative importance of the unequal-mass collisions with respect to the equal-mass collisions. If the unequal-mass collisions of dust with $\taus=\taus_{1}$ and $\taus_{2}$ ($\taus_2<\taus_1<1$) are driven by the radial drift motion, the collision velocity $\Delta v(\taus_1,\taus_2)$ is roughly given by $2(\taus_1-\taus_2)\eta\vk$. Here, we neglect the factor originating from the back reaction from dust to gas, and thus the actual collision velocity is lower. If the mass densities of both species are comparable, the relative importance of the unequal-mass collisions can be measured by ratios of the cross sections and the collision velocities:\footnote{ We examine the relative importance in terms of mass flux to a dust of $\taus=\taus_1$ through collisions rather than in terms of collision probability. The former depends on $\rhod$ while the latter depends on the number density, which is typically assumed to be larger for smaller dust grains.}
\begin{equation}
f\equiv\frac{\pi(a_1+a_2)^2\Delta v(\taus_1,\taus_2)}{4\pi a_1^2\Delta v(\taus_1,\taus_1)},\label{eq:rat_crs_v}
\end{equation}
where $a_1$ and $a_2$ denote the dust sizes corresponding to the Stokes number of $\taus_1$ and $\taus_2$, respectively. Based on Figure \ref{fig:weighted_average}, we can roughly approximate $\Delta v(\taus_1,\taus_1)$ as $\Delta v(\taus_1,\taus_1)\simeq\Delta v_{\mathrm{tur}}(\alpha=10^{-5})$. Assuming $\taus_1\gg\taus_2$ in Equation (\ref{eq:rat_crs_v}), we have
\begin{align}
f&\sim\frac{1}{4}\frac{2\taus_1\eta\vk}{\Delta v_{\mathrm{tur}}(\alpha=10^{-5})}\notag\\
&\sim0.5\left(\frac{\taus_1}{0.01}\right)^{0.5}\left(\frac{\eta\vk/\cs}{0.05}\right).
\end{align}
The factor $f$ should be smaller as we overestimate $\Delta v(\taus_1,\taus_2)$ in overdense regions by a factor of $\sim(1+\epsilon)$ \citep[e.g., see][]{Tanaka2005}. This indicates that growth due to the equal-mass collisions dominates that due to the unequal-mass collisions even if the mass densities are comparable between the two species $(\taus=\taus_1,\taus_2)$. 

 Recent unstratified simulations with multiple dust species showed that dust segregation occurs through the SI-induced turbulence, and large dust grains tend to be abundant in overdense regions \citep[][]{Yang2021,Matthijsse2025}. \citet{Matthijsse2025} showed that the densest regions of different $\taus$ are spatially separated and that local size distributions can have a single peak. Such segregation will promote the growth of large dust grains through equal-mass collisions. Therefore, we expect that unequal-mass collisions might affect insignificantly the evolution of mass-dominating dust grains once the SI-assisted coagulation operates.

 \citet{Bai2010b} showed that the differential drift velocity fits the median collision velocities in most of the runs. However, it is still unclear whether or not such high velocity collisions dominate the dust growth. For example, they mentioned that high-speed collisions take place outside dust clumps, suggesting that the median collision velocity might overestimate characteristic collision velocities for the dust growth. In future work, we will investigate collision velocities and coagulation rates due to unequal-mass collisions in more detail.

 It should be noted that broadening the size distribution may change the efficiency of the SI and thus the equal-mass collision rates. It has been recently highlighted that the SI itself is affected by size distributions. \citet{Krapp2019} showed that the SI with multiple dust species can be significantly slow \citep[see also][]{McNally2021,Zhu2021}. \citet{Zhu2021} showed that the largest dust does not necessarily determine the efficiency of the SI, and smaller dust grains can appreciably change growth rates. Their contribution to the instability is not straightforward since they can change the dominant mode. It will be worth investigating how efficiently small dust grains are swept up regardless of their limited contribution to the growth of mass-dominating large dust grains.

\subsubsection{Stratified disks and three-dimensional simulations}
The present study is based on the unstratified simulations. The previous studies based on stratified simulations show that dust grains are accumulated into azimuthally elongated filaments \cite[e.g.,][]{Johansen2007,Yang2014,Li2018,Schaffer2021}. It was also shown that small dust grains are stirred up by large dust grains that drive clumping around the midplane \citep[e.g.,][]{Bai2010b,Schaffer2018}. Those radial and vertical structures will be important to investigate where and how efficiently the SI-assisted coagulation takes place.

 Besides, vertical shear of the azimuthal velocity in the stratified disk leads to another type of instability called vertically shearing streaming instability (VSSI) \citep[][]{Ishitsu2009,Lin2021}. \citet{Lin2021} found that VSSI dominates over the classic SI in stratified disks. VSSI induces turbulence and stirs dust grains on radial scales of $10^{-3}H$, which is typically smaller than the radial scale of the SI. Thus, VSSI can be another driving source of dust collisions.

Finally, the present study is based on two-dimensional simulations. The clumping efficiency can be different in three-dimensional simulations. For example, the maximum density can be lower in three-dimensional simulations \citep[e.g.,][]{Johansen2007,Bai2010b}. Thus, our analysis might overestimate the coagulation rates. We will address those points in future work.

\section{Conclusion}\label{sec:conclusion}
In this work, we perform unstratified local simulations of the SI and examine how fast dust grains can grow during the moderate clumping. To evaluate the coagulation rates as precisely as possible, we utilize the super-particle trajectories to take into account temporal variation in dust densities and velocities. Our main findings are summarized as follows:
\begin{enumerate}
\item The average coagulation rates are $\sim10^{-2}\Omega-10^{-1}\Omega$ in most of the runs and maximized at $\taus\simeq0.03-0.1$ for a given initial dust-to-gas ratio. The coagulation-rate-weighted average of dust densities is larger than the initial dust density by a factor of a few to ten, meaning that the density enhancement indeed assists the coagulation. 
\item The weighted-average velocity dispersion is on the order of $10^{-3}\cs$ at the highest. Thus, the SI-assisted coagulation cannot be prevented by the fragmentation barrier as suggested in the previous studies \citep[see also][]{Johansen2007supp,Balsara2009,Bai2010b}.
\item The coagulation timescales are orders of magnitude shorter than the drifting timescale. Therefore, the radial drift will not limit the SI-assisted coagulation. For $\taus\lesssim0.1$, the SI-assisted coagulation will produce larger dust grains in the so-called pre-clumping phase.
\item The SI-assisted coagulation appreciably relaxes the conditions for planetesimal formation because (1) it increases $\taus$ and (2) the midplane dust-to-gas ratio $\epmid$ is also enhanced by subsequent settling \citep[see also][]{Ho2024}. According to our simple model, the critical value of the dust-to-gas ratio is $\simeq0.4$ for $\tausini=0.01$, which is only one-tenth of the critical value in the previous study (Figure \ref{fig:epmid_crit}).
\item Our model also suggests that planetesimal formation can be initiated only by moderate clumping that increases the local dust-to-gas ratio from $\epsilon\leq1$ to several (see Figure \ref{fig:weighted_average}). The SI-assisted coagulation and settling realize states where the so-called strong clumping occurs and amplifies the dust density by orders of magnitude.
\end{enumerate}

Our simulations do not treat external gas turbulence, which may enhance collision velocities. However, once the SI develops, such larger-scale turbulence is expected to be inefficient in stirring dust grains \citep[][]{Klahr2020}. Therefore, dust grains will safely grow through the SI-assisted coagulation. Numerical simulations incorporating both the SI and the coagulation is necessary to reveal dust evolution toward planetesimals in more detail \citep[see][]{Ho2024}. Especially, to incorporate the low-velocity collisions driven by the SI is vital since otherwise fragmentation will limit the dust growth regardless of the density enhancement.

\acknowledgments
We thank Min-Kai Lin, Satoshi Okuzumi, Xuening Bai, and Ziyan Xu for valuable and helpful comments. We also thank the anonymous referee for constructive comments that helped us to improve the manuscript. This work was supported by JSPS KAKENHI Grant Nos. 21K20385, 24KJ1041 (R.T.T.), 19K03941 (H.T.).   
\software{Athena \citep{Stone2008}, NumPy \citep{Harris2020}, Matplotlib \citep{Hunter2007}, SciPy \citep{Virtanen2020} }

%




\bibliographystyle{aasjournal}
\bibliography{rttominaga}

%
%
%


\end{document}